%% file: main.tex
\documentclass[conference]{IEEEtran}

\usepackage{comment}
\usepackage[utf8]{inputenc}
\usepackage{xcolor}
\usepackage{hyperref}
\usepackage{xspace}
\usepackage{color}
\usepackage{listings}
\usepackage{acronym}
\usepackage[T1]{fontenc}
\usepackage{multirow}
\usepackage{colortbl}
\usepackage{url}
\usepackage{comment}
\usepackage{colortbl}
\usepackage{lipsum}
\usepackage{multicol}
\usepackage{amsmath}

\usepackage{float}
\usepackage{graphicx}
\usepackage{listings}
\usepackage{amsfonts}
\usepackage{tcolorbox}
\usepackage{wrapfig, blindtext}
\usepackage{longtable}
\usepackage{cancel}
\usepackage{todo}

\usepackage{tabularx}
\usepackage{algorithm}
\usepackage{algpseudocode}
\usepackage{hologo}

\definecolor{lightgray}{rgb}{.9,.9,.9}
\definecolor{darkgray}{rgb}{.4,.4,.4}
\definecolor{darkgreen}{rgb}{0, 0.39, 0.00}
\definecolor{Gray}{gray}{0.7}
\definecolor{codegreen}{rgb}{0,0.6,0}
\definecolor{codegray}{rgb}{0.5,0.5,0.5}
\definecolor{codepurple}{rgb}{0.58,0,0.82}
\definecolor{backcolour}{rgb}{0.95,0.95,0.92}

\lstdefinestyle{mystyle}{
    backgroundcolor=\color{backcolour},
    commentstyle=\color{codegreen},
    keywordstyle=\color{magenta},
    numberstyle=\tiny\color{codegray},
    stringstyle=\color{codepurple},
    basicstyle=\ttfamily\footnotesize,
    breakatwhitespace=false,
    breaklines=true,
    captionpos=b,
    keepspaces=true,
    numbers=left,
    numbersep=5pt,
    showspaces=false,
    showstringspaces=false,
    showtabs=false,
    tabsize=2
}

\include{acronym}

\begin{document}

\IEEEoverridecommandlockouts

\title{Comparison of Fully Homomorphic Encryption and Garbled Circuit Techniques in Privacy-Preserving Machine Learning Inference\\
}

% authors
\author{\IEEEauthorblockN{Kalyan Cheerla}
\IEEEauthorblockA{KalyanCheerla@my.unt.edu}
\and
\IEEEauthorblockN{Lotfi Ben Othmane}
\IEEEauthorblockA{lotfi.benothmane@unt.edu}
\and
\IEEEauthorblockN{Kirill Morozov}
\IEEEauthorblockA{kirill.morozov@unt.edu}
\vspace{-1.in}
\thanks{\hrule\vspace{3mm}}
\thanks{
© 2025 IEEE. Personal use of this material is permitted. Permission from IEEE must be obtained
for all other uses, in any current or future media, including reprinting/republishing this material for
advertising or promotional purposes, creating new collective works, for resale or redistribution to servers or lists,
or reuse of any copyrighted component of this work in other works.\vspace{1mm}}
\thanks{\textit{This is the author's version of the paper accepted for publication in the IEEE Secure Development (SecDev) 2025 Conference, Indianapolis, IN, October 14-16, 2025.}}
}

\maketitle

% abstract
\begin{abstract}
\ac{ML} is making its way into fields such as healthcare, finance, and \ac{NLP}, and concerns over data privacy and model confidentiality continue to grow. \ac{PPML} addresses this challenge by enabling inference on private data without revealing sensitive inputs or proprietary models. Leveraging Secure Computation techniques from Cryptography, two widely studied approaches in this domain are \ac{FHE} and \ac{GC}. This work  presents a comparative evaluation of FHE and GC for secure neural network inference. A two-layer neural network (NN) was implemented using the CKKS scheme from the Microsoft SEAL library (FHE) and the TinyGarble2.0 framework (GC) by IntelLabs. Both implementations are evaluated under the semi-honest threat model, measuring inference output error, round-trip time, peak memory usage, communication overhead, and communication rounds. Results reveal a trade-off: modular GC offers faster execution and lower memory consumption, while FHE supports non-interactive inference.
\end{abstract}


\input{sections/1_introduction}
\input{sections/2_related_works}
\input{sections/3_design_and_implementation}
\input{sections/4_comparitive_analysis}
\input{sections/5_conclusion}
\section*{Acknowledgement}
\noindent The authors used ChatGPT4 to revise Sections 3 and 4 for fixing typos and grammar.

% references/bibiliography
\bibliographystyle{IEEEtran}
\bibliography{references}

% appendix
\input{sections/appendix}

\end{document}

%% file: acronym.tex
\acrodef{ML}{Machine Learning}
\acrodef{FHE}{Fully Homomorphic Encryption}
\acrodef{NLP}{Natural Language Processing}
\acrodef{PPML}{Privacy-preserving Machine Learning}
\acrodef{GC}{Garbled Circuits}
\acrodef{HE}{Homomorphic Encryption}
\acrodef{NN}{Neural Network}
\acrodef{UC}{Universal Circuits}
\acrodef{OT}{Oblivious Transfer}
\acrodef{FSS}{Function Secret Sharing}
\acrodef{SFE}{Secure Function Evaluation}
\acrodef{LUTs}{Lookup Tables}
\acrodef{RTT}{Round-Trip Time}

%% file: sections/1_introduction.tex
\section{Introduction}

AI technologies rapidly spread across sectors, such as healthcare, transportation, and education. In healthcare, for example, \acf{ML} models assist in early disease detection~\cite{wang2020early}, and drug discovery through protein folding~\cite{noe2020machine}. Traditional ML pipelines often require centralized data processing, which can lead to privacy risks in regulated domains such as healthcare. In scenarios where the model provider and the data owner are distinct entities, both parties have conflicting privacy goals: the model provider aims to protect intellectual property (model architecture and parameters), while the data owner seeks to keep their input data private.

\begin{figure}[tbph]
  \centering
    \includegraphics[width=\columnwidth]{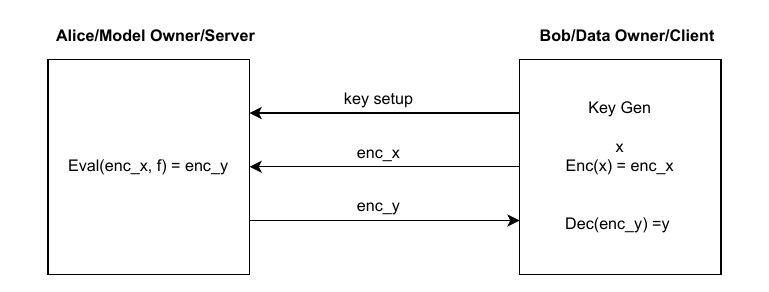}
    \caption[FHE-based inference setup]{FHE-based inference setup: The client encrypts input \(x\) using a public key and sends ciphertext \(enc\_x\), public key, and other required keys except secret key to the server. The server homomorphically evaluates \(f\) to get \(enc\_y\), which the client decrypts using his secret key to recover \(y\).}
  \label{fig:C1_F2}
  \vspace{-0.2 in}
\end{figure}

\begin{figure}[tbph]
  \centering
    \includegraphics[width=\columnwidth]{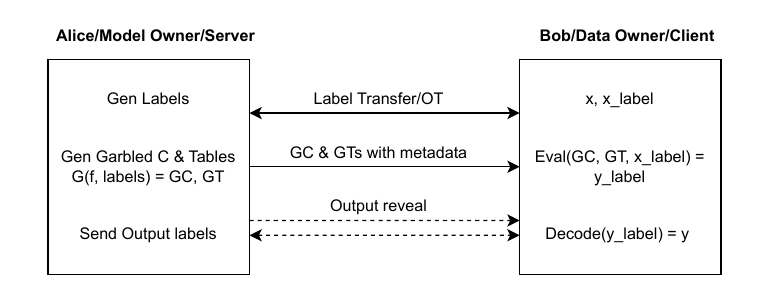}
    \caption[GC-based inference setup]{GC-based inference setup: The server garbles the circuit \(C\) representing \(f\), and both parties engage in an interactive protocol involving label generation, oblivious transfer (OT), transmission of the garbled circuit (\(GC\)) and garbled tables (\(GT\)) with metadata, and secure evaluation (\(Eval\)). Output labels are sent to the client for decoding to obtain the final output \(y\).}
  \label{fig:C1_F3}
  \vspace{-0.2 in}
\end{figure}

Let us consider a typical setting in PPML~\cite{hesamifard2018privacy} that enables \ac{ML} inference while preserving the confidentiality of the ML model and user data in the context of two-party collaboration. Specifically, we have two parties: a \emph{Client} (data owner or Bob), who holds private input data \(x\), and a \emph{Server} (inference model owner or Alice), who owns a pre-trained model \(f(x)\). In our paper, we focus on the case of a fixed-parameter two-layer neural network.  The goal is to perform inference where the Client obtains the prediction \(y = f(x)\) while complying with the following requirements:
\begin{enumerate}
  \item Client input $x$ remains confidential from Server.
  \item Server model $f(x)$ remains confidential from Client.
  \item Correctness of inference $y = f(x)$ is preserved.
\end{enumerate}

The problem of PPML can be addressed using different cryptographic approaches; in this work, we focus on  \ac{FHE} \cite{gentry2009fully} and \ac{GC} \cite{yao1986generate}. Both cryptographic primitives enable building secure computation solutions without exposing raw data~\cite{chandran2023security,mann2023towards}. Both methods support \ac{PPML}, with two inherently different designs that impact their performance, communication overhead, and scalability. These properties are critical to choosing the appropriate method for a specific \ac{PPML} scenario, since each technique exhibits distinct advantages and limitations. GC protocols offer low-latency evaluation and compact representations for Boolean operations. While their interactive nature is often manageable, they do leak the model's structure. FHE schemes, on the other hand, enable non-interactive inference and may support floating-point arithmetic, at the cost of increased computational and memory overhead.

This paper compares the performance trade-offs, and security implications of deploying privacy-preserving neural network inference using \ac{FHE} and \ac{GC} techniques. It presents and discusses the evaluation of these two methods by implementing a two-layer neural network using the following two methods: (1) \ac{FHE} with polynomial approximations for non-linear activation functions to enable computation over encrypted inputs, requiring only a single round of interaction per inference;
as depicted by Figure~\ref{fig:C1_F2};
and (2) A \ac{GC}, where both parties collaborate in a multi-round protocol to evaluate the function securely, as depicted by Figure~\ref{fig:C1_F3}.
The GC protocol allows two parties—commonly referred to as the \textit{garbler} and the \textit{evaluator}—to jointly evaluate a Boolean circuit without revealing their respective private inputs.

This work assumes a semi-honest adversarial model (honest but curious) \cite{mann2023towards}, in both the FHE and GC settings. In this model, both parties in the protocol execution follow the prescribed steps correctly, but they may attempt to infer additional information from the messages they receive. The two primary goals in the private inference setting are:
\begin{itemize}
    \item For Model Owner (Server/Party A/Alice): Holds a trained \ac{ML} model (e.g., weights and biases of the two-layer neural network) and wishes to preserve its confidentiality during inference.
    \item For Data Owner (Client/Party B/Bob): Possesses sensitive input data and wishes to obtain inference results without revealing their data to the server.
\end{itemize}

The key privacy goals are:
\begin{itemize}
    \item The Client should learn only the output of the model evaluation on their input, and nothing about the model's internal parameters.
    \item The Server should not learn the Client's input or the inference result.
\end{itemize}

The study aims to analyze and compare the performance, practicality, and security guarantees of FHE and GC in supporting secure inference tasks. This comparison further seeks to identify conditions under which one approach may be preferable over the other, based on factors such as run time, communication overhead, interactivity, and compatibility with machine learning workloads. The outcomes of this study will offer guidance to researchers and developers looking to adopt the most suitable technique for PPML.

The main contributions of the paper are:
\begin{enumerate}
    \item Evaluation of FHE-based (SEAL-CKKS) \cite{cheon2017homomorphic,ghmsseal} and GC-based (TinyGarble2.0) \cite{hussain2020tinygarble2,ghinteltg2} implementations using the same neural network architecture.
    \item Analysis of the practical trade-offs of FHE and GC protocols in terms of interactivity and approximation effects on model outputs.
\end{enumerate}

The paper is organized as follows. Section~\ref{sec:relatedwork} discusses related work; Section~\ref{sec:Experimentsetup} describes the design and implementation of the experiment used to evaluate the use of \ac{FHE} and \ac{GC} in implementing secure inference of a 2-layer neural network model; Section~\ref{sec:evaluation results} reports about the evaluation results; and Section~\ref{sec:conclusion} concludes the paper.

%% file: sections/2_related_works.tex
\section{Related Work}\label{sec:relatedwork}

A large body of work exists around secure inference using \ac{GC} and \ac{HE}, including frameworks such as SecureML \cite{mohassel2017secureml}, CHET \cite{dathathri2018chet}, and ABY \cite{patra2021aby2}. While these systems focus on protocol-level optimizations and scaling to larger models, they often lack direct, system-level comparisons between different cryptographic paradigms under consistent assumptions. To address this gap, this paper implements a simple two-layer neural network using both TinyGarble2.0 (GC) \cite{ghinteltg2} and Microsoft SEAL (FHE) \cite{ghmsseal}, enabling a practical comparison of latency, communication, memory usage, and interactivity.

Several recent frameworks target secure transformer inference using a variety of \ac{SFE} or hybrid techniques: Sigma \cite{gupta2023sigma}, which leverages \ac{FSS} for secure GPT inference, efficiently implementing non-linearities like Softmax and LayerNorm; Iron~\cite{hao2022iron}, which integrates HE and Secret Sharing to reduce communication in secure BERT inference;
MPCFormer~\cite{li2022mpcformer}, which combines MPC-friendly approximations with knowledge distillation; and CrypTen \cite{knott2021crypten}, SecureGPT \cite{zeng2024securegpt}, and East~\cite{ding2023east}, which optimize the transformer layer protocols, nonlinear components, and secure runtime efficiency.
While these frameworks focus on functionality, optimization, and scaling to larger models, they often overlook fine-grained comparisons between \ac{FHE} and \ac{GC}. Their focus on hybrid MPC approaches and large-scale models leaves a gap in understanding the practical tradeoffs in simpler networks.

Recent works such as \cite{rahman2024benchmarking,zhu2023performance} benchmark FHE libraries or implement secure CNN inference, but they either focus solely on arithmetic microbenchmarks or do not evaluate alternative paradigms like \ac{GC}. Our work complements these by offering a unified, system-level comparison of GC and FHE under a shared inference task and setting. In addition, two recent survey papers \cite{mann2023towards,chandran2023security} provide comprehensive overviews of the PPML field. They categorize secure inference protocols by cryptographic backend, interaction model, and application domain. At the same time, both surveys highlight a lack of consistent benchmarking environments and limited evaluations of system-level performance in minimal secure settings. This paper directly addresses that gap by using a shared model, input, and environment to empirically compare FHE and GC in the end-to-end setting.

%% file: sections/3_design_and_implementation.tex
\section{Experiment Design and Conduct}\label{sec:Experimentsetup}

We selected TinyGarble2.0 for its efficient support for sequential pipelining in \ac{GC}, offering greater flexibility and integration capabilities compared to alternatives such as FLUENT~\cite{gunther2024fluent}. We also selected Microsoft SEAL for \ac{FHE} because it is a mature framework with support for the CKKS scheme enabling approximate arithmetic on real numbers.

\subsection{Neural Network Used}
A simple 2-layer feed-forward neural network with fixed parameters was used for a fair comparison, as shown below:
\[
\mathbf{y} = f(\mathbf{x}) = \text{Sigmoid}\Big(W_2 \cdot \text{ReLU}(W_1 \mathbf{x} + \mathbf{b}_1) + \mathbf{b}_2\Big)
\]

To enable secure inference, the above model's non-linear activations were approximated using low-degree polynomials. ReLU was replaced with a square function, which is commonly selected for its simplicity and ability to preserve non-negativity~\cite{gilad2016cryptonets}. This quadratic form eliminates conditional branching and maintains continuity for polynomial evaluation in \ac{FHE}. Similarly, the sigmoid was approximated using a second-degree polynomial inspired by~\cite{chen2018logistic}, retaining its S-shaped curve with minimal computational overhead. These approximations reduce circuit complexity while preserving core functional properties for efficient encrypted inference.

\begin{table}[tbp]
\caption{Different characteristics across Modes}\label{tab:characteristics_across_modes}
\begin{center}
\begin{tabularx}{\columnwidth}{|l|>{\raggedright\arraybackslash}p{0.08\textwidth}|>{\raggedright\arraybackslash}X|>{\raggedright\arraybackslash}X|}
  \hline
  \textbf{Component} & \textbf{Plaintext} & \textbf{FHE} & \textbf{GC} \\
  \hline
  Input/Output & Plain floating-point vector & Encrypted floating-point vector & Floating-point scaled to fixed-point vector \\
  \hline
  ReLU & \(\max(0, x)\) & \(\approx x^2\) & \(\max(0, x)\) \\
  \hline
  Sigmoid \(\sigma(x)\) & \( \frac{1}{1 + e^{-x}} \) & \(\approx 0.5 + 0.197x - 0.004x^2\) & \(\approx 0.5 + 0.197x - 0.004x^2\) \\
  \hline
\end{tabularx}
\end{center}
\vspace{-.3in}
\end{table}

Table~\ref{tab:characteristics_across_modes} provides the used activation functions and their approximations--See Appendix~\ref{app:appox_activations} for more details. Further details are available in~\cite{cheerla2025thesis}.

\subsection{Fully Homomorphic Encryption-Based Implementation}\label{sec:fhe_implementation}

This section presents the implementation of a two-layer neural network using the leveled CKKS scheme~\cite{cheon2017homomorphic}, as provided by the Microsoft SEAL library~\cite{ghmsseal}. In this setup, the Client (Bob) generates the CKKS keys, encrypts the input vector, and transmits the ciphertext along with the public and evaluation keys to the Server (Alice). The Server, which holds the plaintext model parameters, performs inference directly on the encrypted input and returns the encrypted output to the Client for decryption.
Most of the protocol operates in a non-interactive setting, with interaction required only during the initial key setup, the transfer of encrypted inputs, and the retrieval of encrypted outputs at the end of computation. The Client is not required to remain online during the computation phase--see Appendix \ref{app:secure_inference_algorithms} for the algorithm.

In the SEAL-CKKS scheme setup, the polynomial modulus degree, coefficient modulus chain, and initial scale are key parameters. The polynomial modulus degree determines the system's capacity and security; in this work, we use a degree of $2^{14} =16384$, which supports approximately 438 bits of coefficient modulus under 128-bit security (see Table~\ref{tab:modulus_chain_limits})~\cite{cheon2017homomorphic,ghmsseal}. This degree defines the number of batching slots and directly affects multiplicative depth, computation cost, and memory usage. The coefficient modulus chain comprises a sequence of prime moduli whose bit-lengths sum to the bit budget allowed by the chosen polynomial degree. Each homomorphic multiplication increases the ciphertext scale, while a rescaling operation removes one prime from the chain and reduces the scale. A representative chain such as $[60,40,40,40,30,30]$, totaling 240 bits and supporting up to five multiplication levels with rescaling—well within the 438-bit limit—along with an initial ciphertext scale of $2^{30}$, which provides approximately 6–8 decimal digits of precision per level and balances numerical accuracy with the available noise budget, are used in this work.

\begin{table}[tbp]
\begin{center}
\caption{Polynomial Modulus Degree vs. Modulus Chain Limits\\(128-bit security)}
\label{tab:modulus_chain_limits}
\begin{tabularx}{\columnwidth}{|>{\centering\arraybackslash}p{0.15\textwidth}|>{\centering\arraybackslash}X|>{\centering\arraybackslash}X|}
\hline
\textbf{Polynomial Modulus Degree} & \textbf{Total Bit-Length of Coefficient Modulus Chain} & \textbf{Max Modulus Count} \\ \hline
1024  & 27 bits (not usable for CKKS) & 1  \\ \hline
2048  & 54 bits   & 1  \\ \hline
4096  & 109 bits  & 3  \\ \hline
8192  & 218 bits  & 5  \\ \hline
16384 & 438 bits  & 9  \\ \hline
32768 & 881 bits  & 16 \\ \hline
\end{tabularx}
\end{center}
\vspace{-.3in}
\end{table}

The Client generates the required CKKS keys, including the public key (for encryption), secret key (for decryption), evaluation keys (to enable relinearization after multiplications), and Galois keys (to support slot rotations, which are essential for matrix-vector operations). Once the keys are generated, the Client encodes and encrypts the input vector $x \in \mathbb{R}^3$ into a CKKS ciphertext. Although the ciphertext contains 8192 slots, only the first three are populated with input values, and the rest are padded with zeros.

On the Server side, each row of the first-layer weight matrix $W_1$ is encoded as plaintext. Homomorphic matrix-vector multiplication is implemented by performing element-wise multiplication between the input ciphertext and each plaintext row, followed by a series of slot rotations and additions to compute the inner product. Bias terms $b_1$ are added homomorphically after each multiplication. The output of this layer is then passed through an approximation of the ReLU function, implemented as $z \mapsto z^2$.

The second layer performs a weighted sum using the encoded weights $W_2$, followed by bias addition. The final activation function is approximated using a low-degree polynomial approximation of the sigmoid function--see Table~\ref{tab:characteristics_across_modes}-- which is efficiently evaluable within the CKKS scheme. Once computation is complete, the Server sends the encrypted output back to the Client. The Client then decrypts and decodes the result using the secret key to obtain the final inference output.

\subsection{Garbled Circuits-Based Implementation}\label{sec:gc_implementation}
\begin{figure}[tbp]
\begin{center}
  \includegraphics[width=\columnwidth]{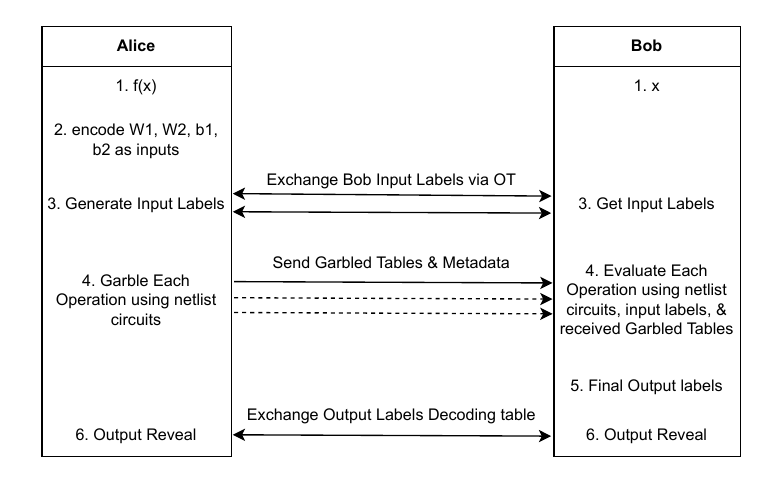}
  \caption[Pipeline view of the GC-based secure inference protocol]{Pipeline view of the GC-based secure inference protocol. This figure depicts the inference pipeline, highlighting party roles, message exchanges, and the sequential evaluation of precompiled circuit netlists.}
  \label{fig:C4_F1}
\end{center}
\vspace{-.3in}
\end{figure}

This section presents the implementation of the same two-layer neural network using the \ac{GC} approach, built with the IntelLabs \texttt{TinyGarble2.0} framework~\cite{hussain2020tinygarble2,ghinteltg2}. Operating in the semi-honest model, this implementation enables secure two-party computation of the neural network through sequential circuit execution, fixed-point arithmetic, and conditional branching for non-linearities. In this setting, the Server (Alice) holds the model parameters and garbles the model as a Boolean circuit, while the Client (Bob) provides the input vector, receives input labels via \ac{OT}~\cite{ghintel_emp_ot}, and jointly evaluates the \ac{GC} using the garbled tables (see Fig~\ref{fig:C4_F1}). The output is reconstructed from the evaluated output labels through an interactive reveal--Appendix \ref{app:secure_inference_algorithms} provides the algorithm. The implementation is available at~\cite{kalyan_snni_fhe_gc_impl}.

In this setup, floating-point values are first scaled (e.g., by 1000) and stored as signed integers. To facilitate secure inference, Alice generates input labels, which are then transmitted to Bob through \ac{OT}. The first layer computation involves matrix multiplication, followed by scale reduction. A bias term $b_1$ is added, and the ReLU activation is applied. The second layer multiplies the ReLU output with the weight matrix $W_2$, applies appropriate scaling, and adds the bias $b_2$. The activation is approximated using a low-degree polynomial: $\sigma(z) \approx 0.5 + 0.197z - 0.004z^2$, which is computed in fixed-point representation with intermediate scaling. Finally, the output of the garbled circuit is revealed.

%% file: sections/4_comparitive_analysis.tex
\section{Analysis of the Evaluation Results}\label{sec:evaluation results}
All experiments were conducted on a Proxmox VM (Ubuntu 24.04, 64-bit) with 8 vCPUs (x86 64-v2 with AES-NI) and 32GB of RAM, hosted on a Intel i5-10400T bare-metal machine. The implementations were compiled with g++13 and CMake3.28.3, with dependencies, including Microsoft SEALv4.1.0, TinyGarble2.0, emp-tool, emp-ot, OpenSSL3.0.13, and Boost1.83.0. Both client and server ran on the same VM to minimize communication overhead and focus on computation time. Each fixed input benchmark was repeated five times, with no variance observed between runs.

\subsection{Round-Trip Time Analysis}
\ac{RTT} measures the total elapsed time from the start of the inference to the final output, which includes input preparation, secure evaluation, and output decoding. This includes key or label generation, input encryption or label transfer, homomorphic or garbled evaluation, and final decryption or output reconstruction. \ac{RTT} was measured in seconds using high-resolution wall-clock timers (\texttt{std::chrono::high\_resolution\_clock}) around the core protocol block. All experiments were performed under identical hardware conditions and averaged over multiple runs to mitigate performance variability.

\begin{figure}[tbp]
\begin{center}
  \includegraphics[width=\columnwidth]{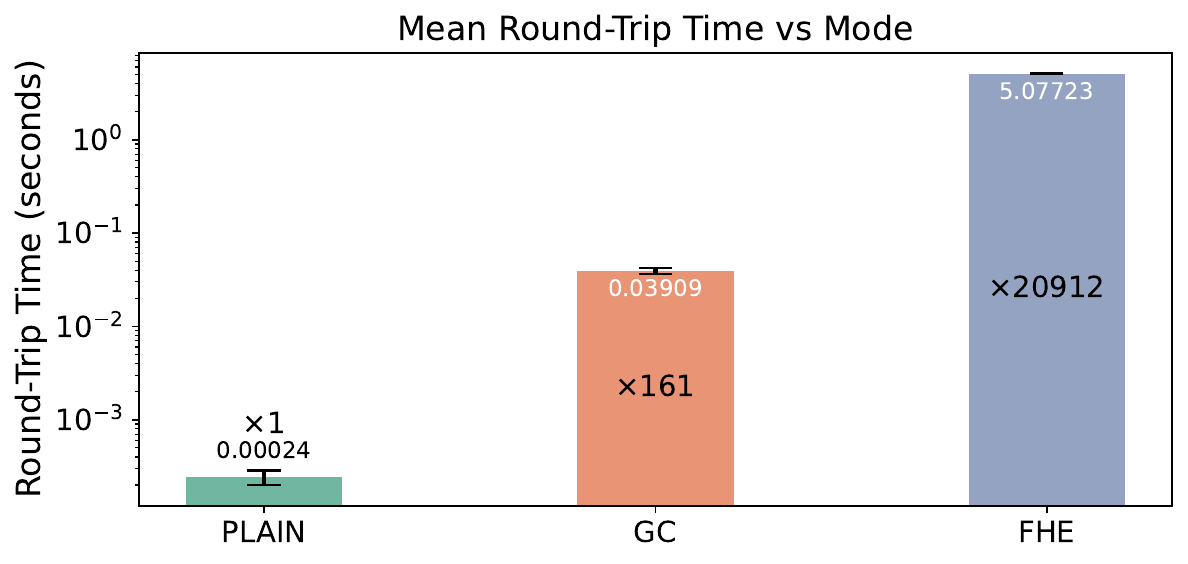}
  \caption[Mean Round-Trip Time vs Modes]{Round-trip time comparison across plain, GC, and FHE modes. The bar chart displays the mean round-trip time for each case on a logarithmic scale. Exact values (in seconds) are shown inside each bar, while slowdown relative to the PLAIN baseline is indicated above the bars as a multiplier. The slowdown is computed as the ratio of each protocol's mean round-trip time to that of the PLAIN baseline, using fixed ordering for fair comparison.}
  \label{fig:C5_F1}
\end{center}
\vspace{-.3in}
\end{figure}

As shown in Fig.~\ref{fig:C5_F1}, computation time varied significantly between the protocols, although remained consistent between Alice and Bob within each. The plaintext baseline--that is, a regular NN implementation without privacy measures--completes in 0.24 ms; it is denoted as PLAIN in Fig.~\ref{fig:C5_F1}. \ac{GC}-based inference is approximately $\times161$ slower than plain text but remains relatively fast, benefitting from TinyGarble2.0’s efficient sequential circuit execution using precompiled netlists. Its runtime is primarily dominated by symmetric-key operations, with communication latency playing a secondary role. In contrast, \ac{FHE}-based inference incurs a substantial $\times20{,}000+$ slowdown, largely due to costly CKKS operations--such as ciphertext multiplications, rescaling, and polynomial approximations of nonlinear functions. The use of a large polynomial modulus (16384) and the lack of batching (i.e., Single Instruction, Multiple Data (SIMD) processing where a single ciphertext encodes multiple plaintext values) also exacerbated it. Because both parties operate on the same machine, network overhead is negligible and the observed \ac{RTT} effectively captures the computation overhead of each protocol. Thus, significant slowdowns are attributed to the different cryptographic workloads imposed by each technique: intensive ciphertext-level arithmetic in \ac{FHE} and gate-level processing in \ac{GC}.

\subsection{Memory Consumption Analysis}
\begin{figure}[htbp]
\begin{center}
  \includegraphics[width=\columnwidth]{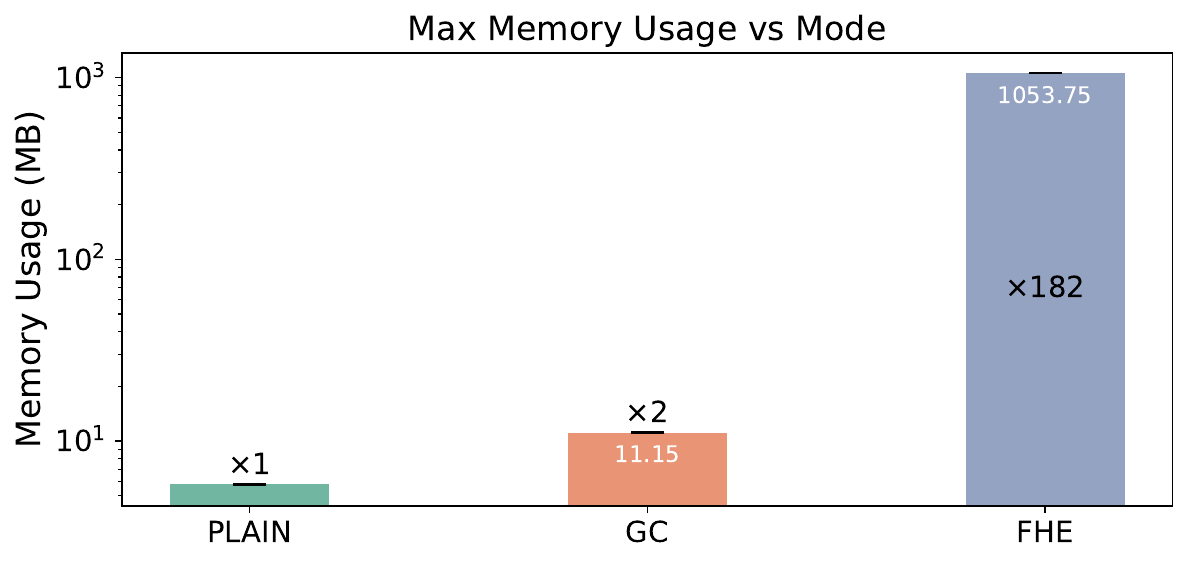}
  \caption[Peak Memory Usage vs Protocol]{Peak memory usage comparison across PLAIN, GC, and FHE protocols. Each bar indicates the mean of the maximum resident set size (MaxRSS) observed across multiple inference runs, where MaxRSS for each run is determined as the higher of the two party-specific memory usages (Alice or Bob). The chart uses a logarithmic scale to visualize large differences in memory footprint across protocols. Relative memory expansion compared to the PLAIN baseline is annotated near or within each bar, computed as the ratio of mean peak memory usage to the PLAIN baseline.}
  \label{fig:C5_F2}
\end{center}
\vspace{-.1in}
\end{figure}

Peak Memory Usage captures the maximum resident set size (MaxRSS) during secure inference, reflecting the peak RAM required by each protocol. This metric is crucial for evaluating feasibility on resource-constrained platforms. Memory usage was profiled programmatically using the \texttt{getrusage()} system call on Linux, with the ru\_maxrss field \cite{man_getrusage} recorded at the end of execution. Measurements include input preparation, secure evaluation, and output decoding.

Based on Fig.~\ref{fig:C5_F2} and the results, \ac{GC}-based inference consumed only $\sim$11.15~MB (server/client), about $\times2$ higher than the plaintext baseline. This efficiency stems from TinyGarble2.0’s sequential and modular evaluation model and shallow intermediate circuits. Memory usage remained stable due to operation-wise clearing of intermediate states. In contrast, \ac{FHE}-based inference consumed 705~MB on the client side and 1053.75~MB on the server. This high usage stems from large polynomial-based ciphertexts, temporary ciphertexts during evaluation, and various keys, all exacerbated by the large polynomial modulus and deep modulus chain.

\subsection{Communication Overhead and Interaction Rounds}
Communication Overhead and Interaction Rounds assess the data transmission and synchronization requirements of each protocol. Two metrics are used: (1) Total bytes transmitted, representing the cumulative volume of data sent by each party, and (2) Communication rounds, defined as logical bidirectional exchanges, incremented whenever the communication direction switches. These metrics were recorded using an instrumented version of the \texttt{NetIO} class from the EMP-tool framework \cite{ghintel_emp_tool,kalyan_emp_tool}, with lightweight counters added to track both data volume and interaction patterns without affecting protocol semantics.

\begin{figure}[htbp]
\begin{center}
  \includegraphics[width=\columnwidth]{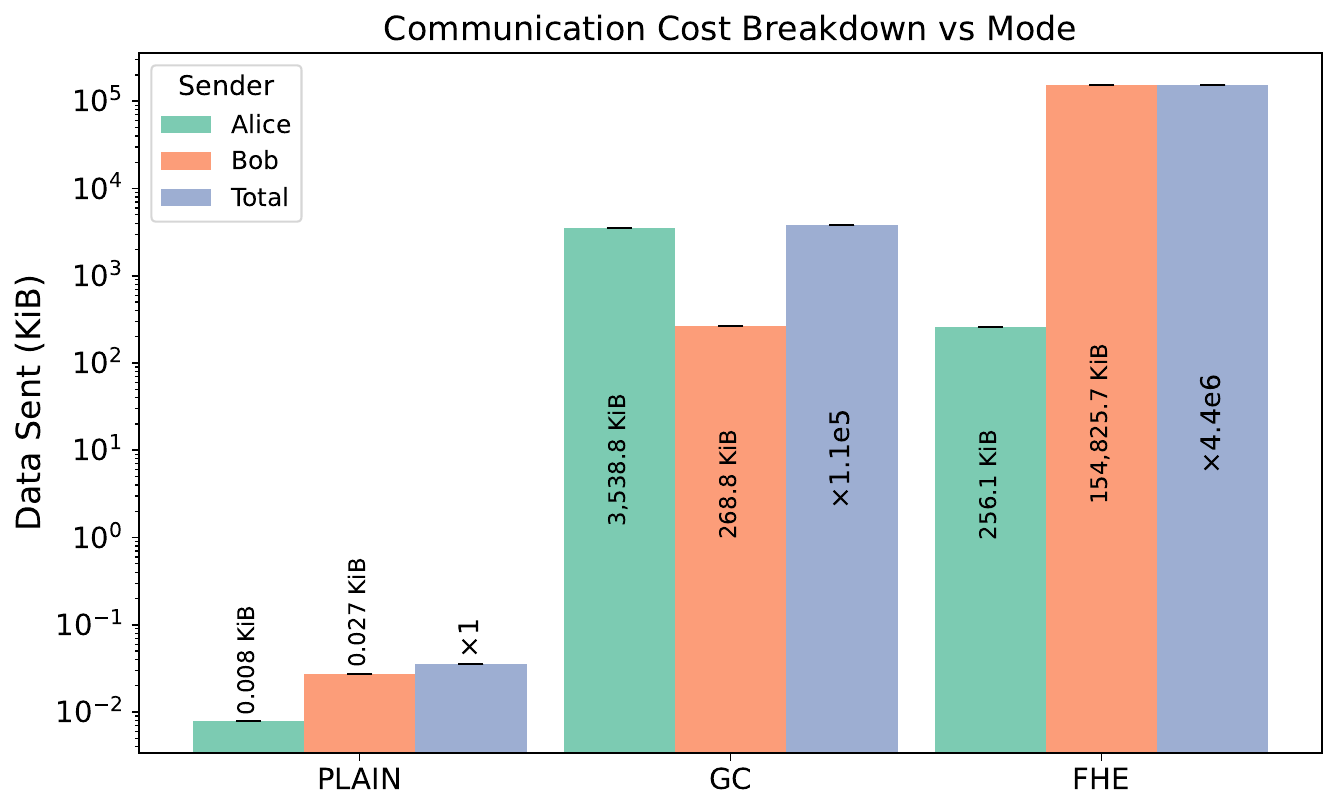}
  \caption[Communication Overhead Breakdown vs Protocol]{Communication overhead comparison across PLAIN, GC, and FHE modes. The chart displays the mean volume of data sent by Alice, Bob, and the combined total for each protocol on a logarithmic scale, computed across multiple inference runs. Individual bars are annotated with human-readable byte values, while the total communication overhead is shown as a scientific multiplier relative to the PLAIN baseline (e.g., $\times1.1 \cdot 10^5$ for GC).}
  \label{fig:C5_F3}
\end{center}
\vspace{-.3in}
\end{figure}

\emph{Communication Overhead}. As shown in Fig.~\ref{fig:C5_F3}, communication volume varied significantly across modes. \ac{GC}-based inference exchanged $\sim$3.76~MiB in total, combining 3.5~MiB from the server and 268~KiB from the client. This includes bandwidth-heavy garbled tables, circuit metadata, and input labels. In contrast, \ac{FHE}-based inference incurred a total cost of around 151.5~MiB, primarily due to the transmission of the input ciphertext and public and evaluation keys. Although this is significantly higher than \ac{GC}’s upper bound, the \ac{FHE} setup can be front-loaded and amortized across multiple inferences when keys are reused. Relative to plaintext inference, \ac{GC} and \ac{FHE} incur $\sim$$\times 1.1 \cdot 10^5$ and $\sim$$\times 4.4 \cdot 10^6$ increases in communication volume, respectively. For \ac{GC}, this disproportionately higher communication cost relative to its $\sim$$\times161$ increase in \ac{RTT} has motivated research efforts focused specifically on reducing communication overhead~\cite{rosulek2021three}.

\emph{Communication Rounds}. \ac{GC} incurred 7 rounds per inference—6 rounds for OT (2 per input) and 1 for output reveal—reflecting its inherently interactive nature. In contrast, \ac{FHE} required only a single round. While \ac{GC} increases interactivity and round complexity, its messages are lightweight, consisting primarily of garbled tables transmitted in one-way interactions that do not add to the communication rounds. In multi-inference scenarios or intermittent networks, \ac{FHE} offers better scalability due to its low interaction and support for batching. Conversely, in bandwidth-constrained settings with fewer inferences, modular \ac{GC} may scale better, especially as model complexity increases.

\begin{figure}[htbp]
\begin{center}
  \includegraphics[width=\columnwidth]{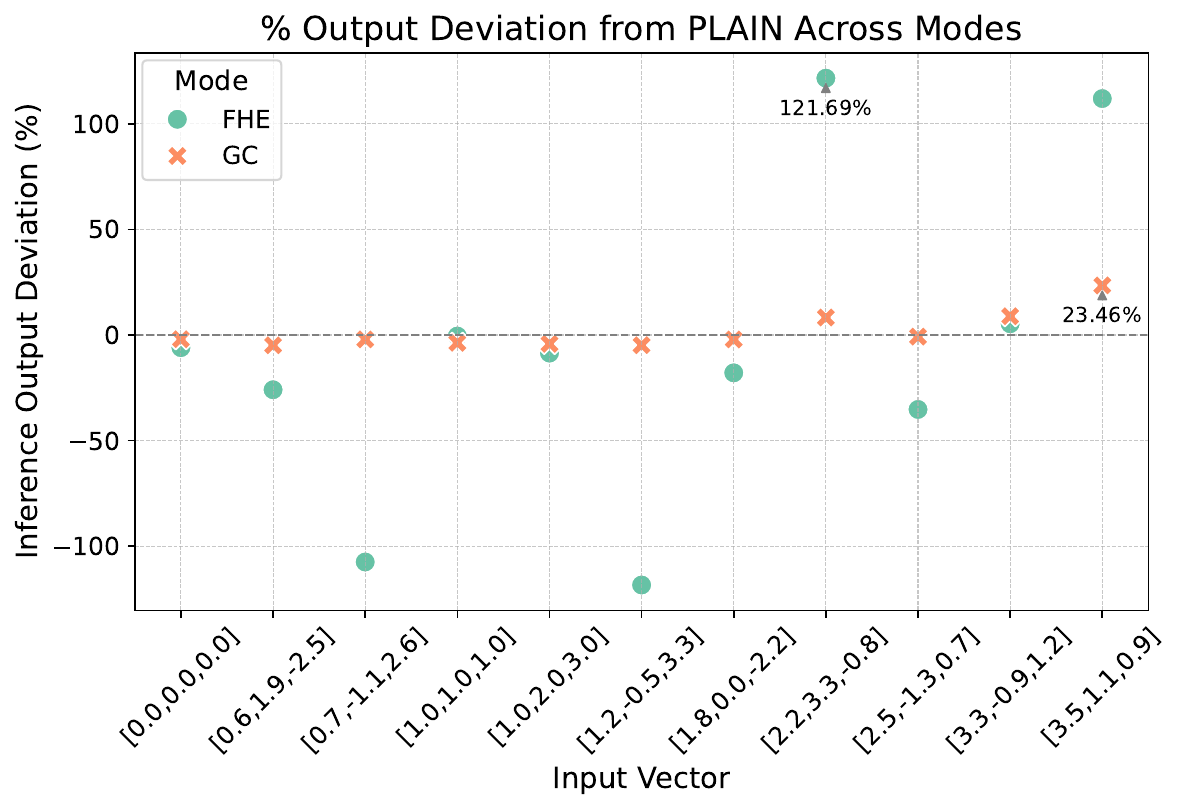}
  \caption[Percentage Output Deviation from PLAIN Across Modes]{Scatter plot showing the percentage deviation of inference outputs from the baseline plain mode for each input vector and protocol.}
  \label{fig:C5_F4}
\end{center}
\vspace{-.3in}
\end{figure}

\subsection{Inference Output Deviation Analysis}
To evaluate the impact of polynomial approximations for non-linear activations in NN-based secure inference, outputs from both approaches were compared against a plaintext baseline. Detailed activation function differences and inference characteristics across modes are presented in Table~\ref{tab:characteristics_across_modes}. Absolute percentage deviations were also computed per input to quantify the effects of these approximations and protocol-specific errors—specifically, fixed-point encoding in \ac{GC} and rescaling operations in \ac{FHE}. The input vectors were randomly chosen to span a diverse set of activation patterns, aiming to stress-test the neural network's non-linearities and expose potential worst-case approximation errors.

As shown in Fig.~\ref{fig:C5_F4}, the outputs in all protocols followed the same trend, though notable deviations were observed. \ac{GC} showed closer alignment with the plaintext baseline, with a worst-case deviation of 23.46\%, primarily due to the second-degree sigmoid approximation and, to a lesser extent, fixed-point scaling. \ac{FHE}, however, exhibited significantly higher variability, with deviations reaching up to 121.69\%. These were mainly caused by the compounded effects of polynomial approximations and rescaling operations. In particular, the non-linearities of activation functions in \ac{FHE} are often lost due to these approximations. While in \ac{GC}, only the sigmoid is approximated, which could be eliminated altogether via \ac{LUTs}, but with increased circuit complexity.

\subsection{Privacy and Scalability Considerations}

\noindent\emph{Privacy Guarantees}. In \ac{FHE}-based inference, both input privacy and full model confidentiality are preserved, as inputs are encrypted and the model remains hidden from the client. In contrast, \ac{GC}-based inference preserves input and parameter privacy but leaks the model structure. This occurs because the Boolean circuit representing the neural network is implicitly shared through the sequence of garbled tables and associated metadata, revealing high-level topology (e.g., number of layers). To mitigate this, \ac{GC} is often combined with \ac{UC}, which supports the evaluation of private functions by embedding logic as control bits within programmable circuits; this improves privacy at the cost of increased computation and communication.

\noindent\emph{Scalability Constraints}. In \ac{FHE}-based inference, scalability is constrained by the multiplicative depth allowed by the chosen modulus chain and polynomial degree (see Table~\ref{tab:modulus_chain_limits}). Without bootstrapping deep networks quickly exhaust noise budgets. In \ac{GC}-based inference, although depth-agnostic in principle, circuit size and label transfer scale linearly with model depth and bitwidth, increasing both computation and communication overhead. While CKKS supports batching via SIMD slots, our experiments used single-input inference; future scalability would require exploiting batching and circuit reuse to support high-throughput scenarios.

\subsection{Generalization to Other Neural Network Architectures}
Our measurements of the inference latency of the evaluated two-layer neural network under CKKS-based FHE show that the initial setup dominates runtime at $\sim$4.8s and computation of each layer
adds negligible cost ($\sim$0.02s), indicating that inference time grows sublinearly with additional layers. This suggests CKKS-based models incur a front-loaded cost, with each layer contributing incrementally. Supporting this, Zhu et al.~\cite{zhu2023performance} evaluated full CNNs over CKKS using SEAL, reporting $\sim$11s latency for a 3-layer CNN on MNIST and $\sim$200s for a deeper CIFAR-10 model, highlighting the compounding effect of deeper activations and convolution-heavy layers. These findings validate our extrapolation and suggest that moderately deeper models, such as 5–10 layer CNNs, would likely incur inference times in the range of 20–60 seconds under comparable parameters.

Our measurements of the GC implementation of the same neural network show that each additional layer in \ac{GC} network increases communication volume linearly--for instance, the first layer resulted in 1.84~MiB of data from Alice, and the addition of a second layer increased the total to 3.54~MiB. This $\sim$1.7~MiB per-layer increment suggests that moderately deeper models of 5–10 layers would incur total one-way communication costs in the range of 8.5–17~MiB. Moreover, unlike \ac{FHE}, where a one-time setup cost (e.g., $\sim$150~MiB of ciphertexts from Bob) enables repeated inferences with low marginal bandwidth ($\sim$1~MiB per inference), \ac{GC} requires re-garbling and full transmission of fresh circuit material for each inference to preserve security. This results in a total communication cost of $\mathcal{O}(n \cdot C)$ for $n$ inferences of a circuit with size $C$, whereas \ac{FHE} achieves $\mathcal{O}(S + n \cdot \varepsilon)$, where $S$ is the initial setup cost and $\varepsilon \ll C$ is the incremental ciphertext overhead per inference. Consequently, \ac{GC} is significantly less bandwidth-efficient than \ac{FHE} in high multi-inference settings. While individual \ac{GC} evaluations remain fast in computation and manageable in bandwidth for LANs, their cumulative cost becomes significant across model depth and inference volume.

%% file: sections/5_conclusion.tex
\section{Conclusion}\label{sec:conclusion}
This work presented a practical comparison of two prominent cryptographic techniques—\ac{FHE} and \ac{GC} for \ac{PPML} inference. By implementing a common two-layer neural network under both paradigms, we evaluated their performance, accuracy, and trade-offs. The results demonstrate that while \ac{GC} offers faster inference and lower memory overhead, it requires interaction and leaks model structure. In contrast, \ac{FHE} ensures full input and model privacy in a non-interactive manner, but at the cost of higher performance overhead. These findings underscore the importance of context-specific considerations when selecting a secure inference method.

Future work will explore extending the current setup to deeper or convolutional models and improving activation function approximations. Hybrid approaches combining \ac{FHE} and \ac{GC}, along with automation for scaling and precision, also represent promising directions.

%% file: sections/appendix.tex
\clearpage
\appendix
\subsection{Secure Inference Protocols for FHE and GC}\label{app:secure_inference_algorithms}
\begin{algorithm}[htbp!]
\caption{FHE-based Secure Inference Protocol}
\label{algo:fhe}
\begin{algorithmic}[1]
\State \textbf{Client (Bob):}
\State Generate CKKS keys (public, secret, relin, galois)
\State Encode and encrypt input vector $x$
\State Send encrypted input and evaluation keys to Server
\State Receive, decrypt and decode final result $y$
\Statex
\State \textbf{Server (Alice):}
\For{each row $w_i$ in $W_1$}
  \State Multiply $enc\_x$ with $w_i$
  \State Rotate and sum to simulate dot product
  \State Add bias $b_i$ and apply ReLU $\approx$ $z^2$
\EndFor
\State Multiply ReLU outputs with $W_2$, add $b_2$
\State Apply sigmoid $\approx 0.5 + 0.197z - 0.004z^2$
\State Send encrypted output to Client
\end{algorithmic}
\end{algorithm}

\begin{algorithm}[htbp!]
\caption{GC-based Secure Inference Protocol}\label{algo:gc}
\begin{algorithmic}[1]
\State \textbf{Server (Alice):}
\State Let model function = $f(x)$ with params $W_1$, $W_2$, $b_1$, $b_2$
\State Initialize $W_1$, $b_1$, $W_2$, $b_2$ as fixed-point integers
\State Encode model values as ALICE inputs \texttt{TG\_int\_init()}
\State Generate input labels for all inputs (client + server)
\For{each operation (add, mult, divscale, matmul, etc.)}
  \State Load the corresponding precompiled circuit netlist
  \State Garble the circuit and generate garbled tables
  \State Send garbled tables and metadata to the client
\EndFor
\Statex
\State \textbf{Client (Bob):}
\State Encode input vector $x$ as fixed-point integers
\State Encode $x$ as BOB inputs using \texttt{TG\_int\_init()}
\State Retrieve input labels via Oblivious Transfer (OT)
\For{each operation}
  \State Receive garbled tables and metadata
  \State Eval circuit via \texttt{sequential\_2pc\_exec\_sh()}
  \State Forward intermediate output labels to next layer
\EndFor
\State Reveal result via \texttt{reveal()} and decode to float
\end{algorithmic}
\end{algorithm}

\newpage
\subsection{Polynomial Approximations of Activation Functions}\label{app:appox_activations}
\begin{figure}[htbp!]
  \centering
    \includegraphics[width=\columnwidth]{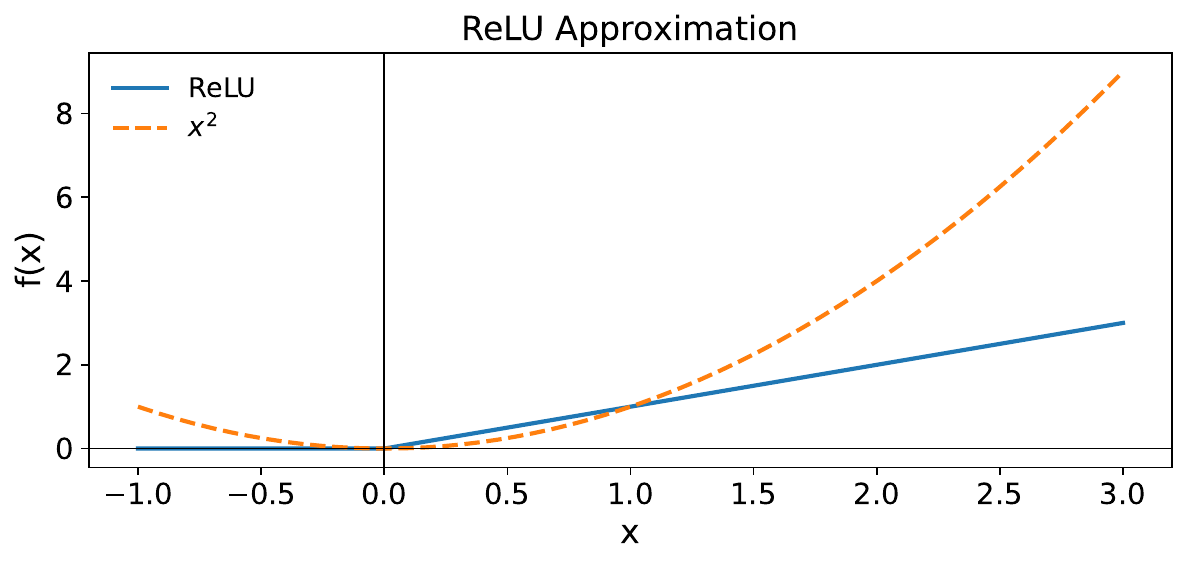}
    \caption[ReLU vs \(x^2\)]{ReLU vs \(x^2\): The approximation preserves non-negativity but diverges significantly for large inputs.}
  \label{fig:A1_F1}
\end{figure}
\begin{figure}[htbp!]
  \centering
    \includegraphics[width=\columnwidth]{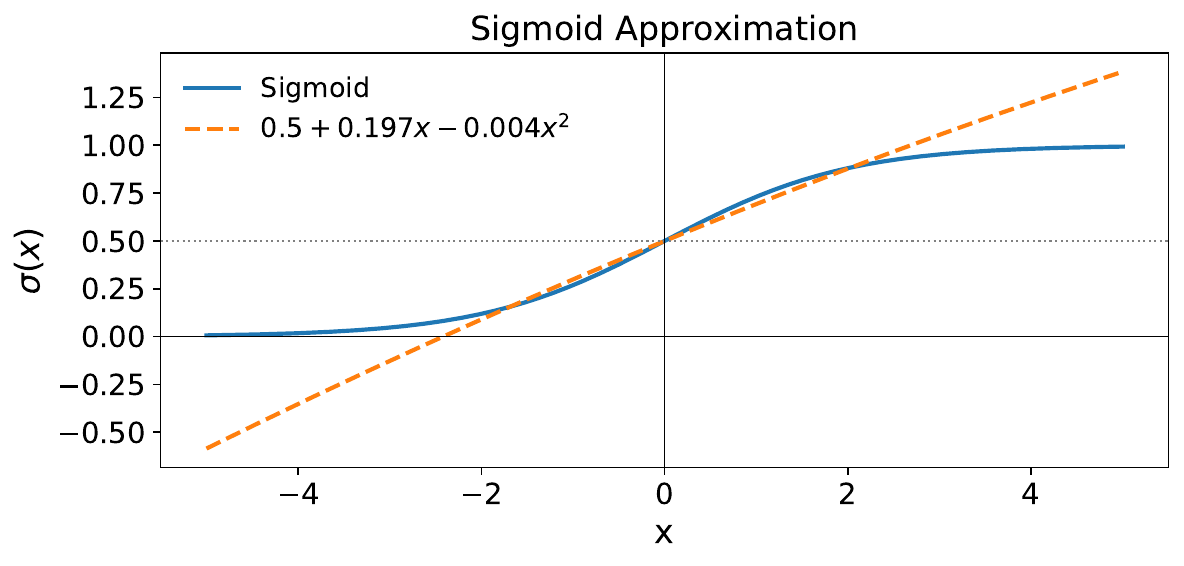}
    \caption[Sigmoid vs \( 0.5 + 0.197x - 0.004x^2 \)]{Sigmoid vs \( 0.5 + 0.197x - 0.004x^2 \): The approximation retains the sigmoid-like shape for small inputs but diverges for large positive values.}
  \label{fig:A1_F2}
\end{figure}

%% file: main.bbl
% Generated by IEEEtran.bst, version: 1.12 (2007/01/11)
\begin{thebibliography}{10}
\providecommand{\url}[1]{#1}
\csname url@samestyle\endcsname
\providecommand{\newblock}{\relax}
\providecommand{\bibinfo}[2]{#2}
\providecommand{\BIBentrySTDinterwordspacing}{\spaceskip=0pt\relax}
\providecommand{\BIBentryALTinterwordstretchfactor}{4}
\providecommand{\BIBentryALTinterwordspacing}{\spaceskip=\fontdimen2\font plus
\BIBentryALTinterwordstretchfactor\fontdimen3\font minus
  \fontdimen4\font\relax}
\providecommand{\BIBforeignlanguage}[2]{{%
\expandafter\ifx\csname l@#1\endcsname\relax
\typeout{** WARNING: IEEEtran.bst: No hyphenation pattern has been}%
\typeout{** loaded for the language `#1'. Using the pattern for}%
\typeout{** the default language instead.}%
\else
\language=\csname l@#1\endcsname
\fi
#2}}
\providecommand{\BIBdecl}{\relax}
\BIBdecl

\bibitem{wang2020early}
W.~Wang, J.~Lee, F.~Harrou, and Y.~Sun, ``Early detection of parkinson’s
  disease using deep learning and machine learning,'' \emph{IEEE Access},
  vol.~8, pp. 147\,635--147\,646, 2020.

\bibitem{noe2020machine}
F.~No{\'e}, G.~De~Fabritiis, and C.~Clementi, ``Machine learning for protein
  folding and dynamics,'' \emph{Current opinion in structural biology},
  vol.~60, pp. 77--84, 2020.

\bibitem{hesamifard2018privacy}
E.~Hesamifard, H.~Takabi, M.~Ghasemi, and R.~N. Wright, ``Privacy-preserving
  machine learning as a service,'' \emph{Proceedings on Privacy Enhancing
  Technologies}, 2018.

\bibitem{gentry2009fully}
C.~Gentry, \emph{A fully homomorphic encryption scheme}.\hskip 1em plus 0.5em
  minus 0.4em\relax Stanford university, 2009.

\bibitem{yao1986generate}
A.~C.-C. Yao, ``How to generate and exchange secrets,'' in \emph{27th Annual
  Symposium on Foundations of Computer Science (sfcs 1986)}, 1986, pp.
  162--167.

\bibitem{chandran2023security}
N.~Chandran, ``Security and privacy in machine learning,'' in
  \emph{International Conference on Information Systems Security}.\hskip 1em
  plus 0.5em minus 0.4em\relax Springer, 2023, pp. 229--248.

\bibitem{mann2023towards}
Z.~{\'A}. Mann, C.~Weinert, D.~Chabal, and J.~W. Bos, ``Towards practical
  secure neural network inference: the journey so far and the road ahead,''
  \emph{ACM Computing Surveys}, vol.~56, no.~5, pp. 1--37, 2023.

\bibitem{cheon2017homomorphic}
J.~H. Cheon, A.~Kim, M.~Kim, and Y.~Song, ``Homomorphic encryption for
  arithmetic of approximate numbers,'' in \emph{Advances in
  cryptology--ASIACRYPT 2017: 23rd international conference on the theory and
  applications of cryptology and information security, Hong kong, China,
  December 3-7, 2017, proceedings, part i 23}.\hskip 1em plus 0.5em minus
  0.4em\relax Springer, 2017, pp. 409--437.

\bibitem{ghmsseal}
``{M}icrosoft {SEAL} (release 4.1),'' \url{https://github.com/Microsoft/SEAL},
  Jan. 2023, microsoft Research, Redmond, WA.

\bibitem{hussain2020tinygarble2}
S.~Hussain, B.~Li, F.~Koushanfar, and R.~Cammarota, ``Tinygarble2: smart,
  efficient, and scalable yao's garble circuit,'' in \emph{Proceedings of the
  2020 Workshop on Privacy-Preserving Machine Learning in Practice}, 2020, pp.
  65--67.

\bibitem{ghinteltg2}
{Intel Labs}, ``{TinyGarble2.0: High-Performance Sequential Garbled
  Circuits},'' \url{https://github.com/IntelLabs/TinyGarble2.0}, 2023,
  accessed: 2025-04-19.

\bibitem{mohassel2017secureml}
P.~Mohassel and Y.~Zhang, ``Secureml: A system for scalable privacy-preserving
  machine learning,'' in \emph{2017 IEEE symposium on security and privacy
  (SP)}.\hskip 1em plus 0.5em minus 0.4em\relax IEEE, 2017, pp. 19--38.

\bibitem{dathathri2018chet}
R.~Dathathri, O.~Saarikivi, H.~Chen, K.~Laine, K.~Lauter, S.~Maleki,
  M.~Musuvathi, and T.~Mytkowicz, ``Chet: compiler and runtime for homomorphic
  evaluation of tensor programs,'' \emph{arXiv preprint arXiv:1810.00845},
  2018.

\bibitem{patra2021aby2}
A.~Patra, T.~Schneider, A.~Suresh, and H.~Yalame, ``Aby2. 0: New efficient
  primitives for stpc with applications to privacy in machine learning,'' in
  \emph{NeurIPS 2021 Workshop Privacy in Machine Learning}, 2021.

\bibitem{gupta2023sigma}
K.~Gupta, N.~Jawalkar, A.~Mukherjee, N.~Chandran, D.~Gupta, A.~Panwar, and
  R.~Sharma, ``Sigma: Secure gpt inference with function secret sharing,''
  \emph{Cryptology ePrint Archive}, 2023.

\bibitem{hao2022iron}
M.~Hao, H.~Li, H.~Chen, P.~Xing, G.~Xu, and T.~Zhang, ``Iron: Private inference
  on transformers,'' \emph{Advances in neural information processing systems},
  vol.~35, pp. 15\,718--15\,731, 2022.

\bibitem{li2022mpcformer}
D.~Li, R.~Shao, H.~Wang, H.~Guo, E.~P. Xing, and H.~Zhang, ``Mpcformer: fast,
  performant and private transformer inference with mpc,'' \emph{arXiv preprint
  arXiv:2211.01452}, 2022.

\bibitem{knott2021crypten}
\BIBentryALTinterwordspacing
B.~Knott, S.~Venkataraman, A.~Y. Hannun, S.~Sengupta, M.~Ibrahim, and
  L.~van~der Maaten, ``Crypten: Secure multi-party computation meets machine
  learning,'' \emph{CoRR}, vol. abs/2109.00984, 2021. [Online]. Available:
  \url{https://arxiv.org/abs/2109.00984}
\BIBentrySTDinterwordspacing

\bibitem{zeng2024securegpt}
C.~Zeng, D.~He, Q.~Feng, X.~Yang, and Q.~Luo, ``Securegpt: A framework for
  multi-party privacy-preserving transformer inference in gpt,'' \emph{IEEE
  Transactions on Information Forensics and Security}, 2024.

\bibitem{ding2023east}
Y.~Ding, H.~Guo, Y.~Guan, W.~Liu, J.~Huo, Z.~Guan, and X.~Zhang, ``East:
  Efficient and accurate secure transformer framework for inference,''
  \emph{arXiv preprint arXiv:2308.09923}, 2023.

\bibitem{rahman2024benchmarking}
T.~Rahman, A.~M. I.~M. Osmani, M.~S. Rahman, M.~M.~A. Shibly, and S.~Islam,
  ``Benchmarking fully homomorphic encryption libraries in iot devices,'' in
  \emph{Proceedings of the 11th International Conference on Networking,
  Systems, and Security}, 2024, pp. 16--23.

\bibitem{zhu2023performance}
H.~Zhu, T.~Suzuki, and H.~Yamana, ``Performance comparison of homomorphic
  encrypted convolutional neural network inference among helib, microsoft seal
  and openfhe,'' in \emph{2023 IEEE Asia-Pacific Conference on Computer Science
  and Data Engineering (CSDE)}.\hskip 1em plus 0.5em minus 0.4em\relax IEEE,
  2023, pp. 1--7.

\bibitem{gunther2024fluent}
D.~G{\"u}nther, J.~Schmidt, T.~Schneider, and H.~Yalame, ``Fluent: A tool for
  efficient mixed-protocol semi-private function evaluation,'' \emph{Cryptology
  ePrint Archive}, 2024.

\bibitem{gilad2016cryptonets}
R.~Gilad-Bachrach, N.~Dowlin, K.~Laine, K.~Lauter, M.~Naehrig, and J.~Wernsing,
  ``Cryptonets: Applying neural networks to encrypted data with high throughput
  and accuracy,'' in \emph{International conference on machine learning}.\hskip
  1em plus 0.5em minus 0.4em\relax PMLR, 2016, pp. 201--210.

\bibitem{chen2018logistic}
H.~Chen, R.~Gilad-Bachrach, K.~Han, Z.~Huang, A.~Jalali, K.~Laine, and
  K.~Lauter, ``Logistic regression over encrypted data from fully homomorphic
  encryption,'' \emph{BMC medical genomics}, vol.~11, pp. 3--12, 2018.

\bibitem{cheerla2025thesis}
K.~Cheerla, ``Comparison of fully homomorphic encryption and garbled circuits
  approaches in privacy-preserving machine learning,'' Master's thesis,
  University of North Texas, 2025.

\bibitem{ghintel_emp_ot}
{Intel Labs}, ``{emp-ot: Oblivious Transfer Protocols for EMP-Toolkit},''
  \url{https://github.com/IntelLabs/emp-ot}, 2016, accessed: 2025-04-19.

\bibitem{kalyan_snni_fhe_gc_impl}
K.~Cheerla, ``{Secure Neural Network Inference using FHE and GC: Implementation
  and Benchmarking},'' \url{https://github.com/kalyancheerla/snni-fhe-gc},
  2025, contains complete two-layer neural network implementations using
  Microsoft SEAL (CKKS) and TinyGarble2.0, with benchmarking support for
  runtime, memory, and communication. Accessed: 2025-04-23.

\bibitem{man_getrusage}
M.~Kerrisk, ``getrusage(2) - linux manual page,''
  \url{https://man7.org/linux/man-pages/man2/getrusage.2.html}, 2024, accessed:
  2025-04-19.

\bibitem{ghintel_emp_tool}
{Intel Labs}, ``{emp-tool: Efficient Secure Computation Library},''
  \url{https://github.com/IntelLabs/emp-tool}, 2016, accessed: 2025-04-19.

\bibitem{kalyan_emp_tool}
K.~Cheerla, ``{Modified EMP-Tool with Communication Statistics Support},''
  \url{https://github.com/kalyancheerla/emp-tool}, 2025, forked from IntelLabs
  EMP-Toolkit with added instrumentation for communication tracking. Accessed:
  2025-04-19.

\bibitem{rosulek2021three}
M.~Rosulek and L.~Roy, ``Three halves make a whole? beating the half-gates
  lower bound for garbled circuits,'' in \emph{Annual International Cryptology
  Conference}.\hskip 1em plus 0.5em minus 0.4em\relax Springer, 2021, pp.
  94--124.

\end{thebibliography}
